# Synthesis of nanoparticles from carboxymethyl cellulose using one-pot hydrothermal carbonization for Drug Entrapment Studies

**Mohaddeseh Sharifi and S.Hajir Bahrami[1]**

*Textile Engineering Department, Amirkabir University of Technology, Tehran, Iran*

**Abstract**

Porous nanomaterials have recently attracted a lot of attention due to various properties and potential applications. In this study, carbon nanoparticles (CNPs) were synthesized by the one-pot hydrothermal carbonization (HTC) using carboxymethyl cellulose (CMC). Urea was used as the nitrogen source for carbonization. The presence of urea in CMC solution for carbonization resulted in CNPsu reduction in the diameter of particles from 4 μm to 1μm. Activation process at high temperature for both the above samples resulted in nanoparticles with diameter of 51 nm and 31 nm, respectively. The positive effect of presence urea and its activation generated different functional groups including C-N, N-H, and C≡N with increasing aromatic rings that probably may help entrapment of drugs into them. On the other hand, activation CNPsu (ACNPsu) has the most aromatic rings with the lowest hydroxyl groups with 84.66% carbon and 12.29% oxygen in its structures.  ACNPs, and ACNPsu exhibited a type I isotherm indicating microporous materials with a high surface area about 552.9 $m^2$/g and 351.01 $m^2$/g, respectively. The high surface area was characteristic of activated carbons with their high adsorption capacity. Thus, the synthesized materials were characterized using SEM, TEM, DLS, BET, FTIR, HNMR, and TGA techniques. Finally, the encapsulation of clindamycin drug (CD) with positive charge in different types of NPs with negative charge was investigated for drug delivery in biomedical engineering applications.

---

[1] Corresponding author.

E-mail address: hajirbahrami@gmail.com.  (S.H. Bahrami)

1. Introduction

In recent years, nanotechnology has attracted considerable attention in a wide range of research areas, focusing on the creation of nanoscale materials using various techniques. Nanoparticles, with diameters ranging from 1 to 100 nm, generate large surface-to-volume ratios with a wide variety of shapes and exhibit distinct and adjustable properties that set them apart from their larger, bulk counterparts [1, 2]. Carbon nanomaterials indicate distinctive properties that are influenced by various characteristic parameters, such as shape, size, chemical composition, surface characteristics, good biocompatibility, non-toxicity, and high mechanical/thermal properties. These factors are promising parameters for applications in electrochemical biosensing, materials science, nanofabrication energy, biomedicine, gene delivery, and drug delivery. Also, other applications of carbon nanomaterials are related to cellular uptake, biodistribution, and interactions that can penetrate cells more easily. So, smaller nanoparticles with their surface charge can penetrate cells more easily and interact with cell membranes. A comprehensive understanding of these parameters is crucial for the strategic design of carbon nanomaterials in biomedical applications, aiming to improve safety and efficacy in nanomedicine [2, 3].

In the last twenty years, carbohydrates or carbohydrate-rich biomass such as carboxymethyl cellulose (CMC), waste cellulose diacetate (CDA), starch, chitosan, lignin, hemicellulose, and cellulose with their renewable, highly abundant precursors, cheap, and green properties increasingly are used to produce carbon nanoparticles (CNPs). CNPs indicate excellent features, including easy preparation and surface modification, excellent photostability, resistance to light bleaching, tunable fluorescence emission, distinctive optical properties, and environment-friendly. To synthesize CNPs, among different techniques such as arc discharged soot, laser ablation carbon materials, electrochemical method, and microwave-assisted polyol, the hydrothermal carbonization (HTC) of biomass is an innovative, efficient, low-cost, sustainable, green technique, and eco-friendly method for transforming biomass into more valuable carbon-based materials with a promising route for broad potential applications. The reactions take place in a closed system at moderate temperatures (120-280°C) and pressure is created naturally. This reaction prevents harsh conditions such as strong acids, high temperatures, and long reaction times [4]. Cheng et al. used lignin as a three-dimensional network of aromatic polymers with a stable phenolic structure for the synthesis of carbon particles by the HTC method. Carbon spheres produced were zero-dimensional carbon materials types with homogeneous morphology, and controllable size. Also, the results demonstrated that different parameters such as biomass feedstock, temperaute, residence time, and pressure in HTC method in synthesizing carbon particles were affected the structure of carbon materials [5]. Also, Barbero et al. synthesized CNPs from glucose by HTC method as a sustainable and green technique to produce carbonaceous nanomaterials. The results showed that synthetic conditions had influenced the particle size. Thus, increasing the synthesis time or the precursor concentration augmented the mean CNPs size [6]. Moreover, Jin et al. used cellulose to fabricate novel carbon dots (CDs) with high quantum yield. CDs had the cationic imidazolium groups on their surface to exhibit excellent fluorescence stability. Also, CDs with uniform 0D nanoparticles had average diameter of 5.57 nm with 80.82% carbon, 11.01% oxygen, 7.67% nitrogen that belonged to amorphous structure [7].

CMC is a naturally sourced polymer and a cellulose derivative known for its water solubility. The presence of hydrophilic groups ($CH_2COO^-$) enhances its capacity to form hydrogen bonds, which adds to its advantageous characteristics. CMC is a semi-crystalline, water-soluble, non-toxicity, excellent biodegradability, high biocompatibility, and environmentally friendly polymer, featuring hydrophilic carboxyl groups along its molecular chain. However, in its pure, single-phase form, CMC shows limited miscibility and reduced chemical stability under specific conditions [8, 9].

To modifiy carbon, nanomaterials N- doping as an efficient method during their synthesis through nitrogen or oxygen-rich precursors or activation of carbon nanomaterials to generate N- containing group in their surface is used. N-doping can influence the structural arrangement, surface functional groups, stabilities, and thermal properties, which further expands the range of applications. The nitrogen source and the doping methods greatly influence the surface groups, the N content, and the structure. Wu et al. synthesized conductive carbon spheres (CSs) and water-soluble fluorescent carbon nano-dots (CNDs) by the one-pot HTC method of carboxymethylcellulose (CMC) and urea as the nitrogen source. The presence of urea in the reaction resulted in creation of nitrogen-containing functional groups such as -$NH_2$- and -NH- groups [4]. Also, Zhao et al. reported the synthesis of N-doped carbon dots (N-CDs) from waste cellulose diacetate (CDA) via one-pot HTC method in aqueous solution. N-CDs exhibited functional groups such as N-H, and C-N in their structure because of the N-doping method for surface modification of CDs [10, 11]. In another study, Shen et al. focused on investigating the effect of urea/cellulose ratio, temperature, and reaction time to synthesize of N-doped CDs by one-pot HTC method. Nitrogen doping on the CDs can affect on the electric properties, surface, and local chemical relativities of CDs. The results showed N-CDs had a high photoluminescent quantum yield of about 21.7% for bioimaging reagent application [12].

In this study, carbon nanoparticles (CNPs) were synthesized by one-pot HTC method using of carboxymethyl cellulose (CMC) in a stainless steel autoclave in an oven at 210°C for 12 h. Also, urea as a nitrogen source was incorporated with CMC solution to obtain carbon nanoparticles/urea (CNPsu). On the other hand, activation of CNPs, and CNPsu in high temperature at 900°C in an nitrogen furnace for 2 hours resulted in generation of activated carbon nanoparticles (ACNPs), and activated carbon nanoparticles/urea (ACNPsu), respectively. The functional groups of solid carbon spheres , the chemical structures of them, the inner structure and the elemental composition were investigated by FTIR, HNMR, and EDX, respectively. Moreover, the morphology and diameter of NPs were observed using SEM, and TEM. Also, pore size and surface area of them were evaluated by DLS, and BET, respectively. Finally, the percentage encapsulation efficiency of clindamycin drug (CD) into different types of NPs was studied during 24.

## 2. Materials and methods

### 2.1. Materials

Carboxymethyl cellulose (CMC) with an average molecular weight of 250,000 and urea were purchased from Sigma Aldrich chemical company, and Phosphate buffer saline (PBS, PH= 7.4) was prepared from Roman Industrial Company (USA). Also, clindamycin drugs were purchased from Mofid pharmaceutical co.

### 2.2. Synthesis of nanoparticles (NPs)

Carboxymethyl cellulose (CMC) (1.5 g) with urea (0.2 g) and/or without urea were dissolved in distilled water (80 mL) for 24 h, separately. The solutions were added to a stainless steel autoclave in an oven at 210° C for 12 h and then allowed to cool at room temperature. The black solid products were collected by centrifugation at 10000 rpm and washed with distilled water and pure ethanol until the solution was clear. The samples were dried under vacuum at 80°C overnight to obtain CNPs, and CNPsu. To activate, CNPs and CNPsu samples were placed in an nitrogen furnace at 900° C for 2 hours and were allowed to cool at room temperature. Finally, the products were washed with boiling water to remove the excess urea decomposition products to fabricate ACNPs, and ACNPsu, respectively [4].

2.3. Characterization of NPs

The size and morphology of CNPs, CNPsu, ACNPs, and ACNPsu were characterized using scanning electron microscopy (SEM) and Transmission Electron Microscopy (TEM). In SEM, four different types of NPs were coated with gold at an accelerating voltage of 15 kV, and images were made in various magnifications. Also, TEM as an invaluable tool has been used to dip a copper grid on the PEM-100 electron microscopy at room temperature at a voltage of 200 kV. The particle size distribution of the aqueous dispersions in viscosity of 0.88 cP was measured by a dynamic light scattering (DLS) device with the Brownian motion of the NPs, and the intensity of scattered laser light fluctuates with time. So, the distribution of particle sizes with protein disulfide Isomerase (PDI) and diameter of them were studied by the SBL algorithm. To investigate the porous structure of 4 types of NPs was used of $N_2$ adsorption-desorption isotherm at 77.35 k with an ASAP 2020 instrument (Mircometritics, USA). The Brunauer-Emmett-Teller (BET) surface area was calculated from the $N_2$ adsorption isotherms in a Quantachrome surface area analyzer (model Autosorb 1) using the BJH adsorption method. Also, the total pore volume and total pore size were examined at a relative pressure of 0.01-0.1. The functional groups of CNPs, CNPsu, ACNPs, and ACNPsu were recorded by the Fourier transform infrared (FTIR) instrument from 4000 cm$^{-1}$ to 400 cm$^{-1}$ at a resolution of 4 cm$^{-1}$. The chemical structure of 4 types of NPs was evaluated using Porton Nuclear Magnetic Resonance (HNMR). The inner structure and the elemental composition in the nanometer ranges were analyzed using Energy-dispersive X-ray Spectroscopy (EDS). Thermogravimetric analysis (TGA) was performed using a Pyris1 TGA apparatus (TA Instrument, USA) at a heating rate of 10 °C min$^{-1}$ under an $N_2$ flow of 20 ml min$^{-1}$. Finally, the best nanoparticles in morphology, pore size, surface area, and chemical structure were selected to incorporate with the drug to investigate encapsulation efficiency of CD into NPs for biomedical applications. Finally, the charge and electrophoretic mobility of ACNPsu were investigated by zeta meter (Zeta-Meter System 3.0+, Zeta Meter Inc., and USA) to load drugs into them.

2.4. Encapsulation efficiency of the drug into NPs

To evaluate the amount of drug in nanoparticles, CD in different concentrations (0.0005, 0.001, and 0.002) g/mL was loaded into NPs (0.001) g/mL in PBS solution. The CD was incorporated into four types of NPs and capsules were produced that were including CD/CNPs, CD/CNPsu, CD/ACNPs, and CD/ACNPsu. The samples were stirred for 2 hours and were ultrasonic for 10 min. Then, samples were placed in a shaking incubator at 37° C and were centrifuged at 4000 rpm. A PBS solution of about 3 mL was removed from each tube to record the UV-vis absorption spectra at 203 nm. The amount of entrapped drug was calculated with a standard equation from the calibration curve. The drug entrapment into NPs includes Eq. (1):

Drug entrapment (%) = $\frac{Amount\ of\ entrapped\ drug}{Theoretical\ drug\ content} \times 100$     (1)

## 3. Results and discussions

### 3.1. Fourier transform infrared (FTIR)

Chemical structure and functional groups of NPs were investigated by FTIR analysis. Fig. 1 shows functional groups of CMC, CNPs, CNPsu, ACNPs, and ACNPsu in different absorption bonds. CMC solution to synthesis CNPs has peaks including 576 cm$^{-1}$, 1021 -1417 cm$^{-1}$, 1634 cm$^{-1}$, 2925 cm$^{-1}$, and 3397 cm$^{-1}$ that belong to C-H bending vibrations, asymmetric (C-O-C) ether stretching, carbonyl (C=O) stretching, C-H stretching vibration, and hydroxyl (O-H) groups, respectively. Also, CNPs that were synthesized from CMC solution have functional groups including C-H stretching vibration, C-O-C groups, C=C, C=O, CH$_3$ stretching vibrations, and O-H groups in wave numbers of 799 cm$^{-1}$, 1289-1379 cm$^{-1}$, 1438 cm$^{-1}$, 1608 cm$^{-1}$, 1701 cm$^{-1}$, 2927 cm$^{-1}$, and 3399 cm$^{-1}$, respectively (Fig. 1(a)). Moreover, adding urea to the CMC solution to produce CNPsu was caused to generate peaks with N groups. So, CNPsu have a broad absorption around 3364 cm$^{-1}$ is assigned to N-H stretching vibrations. Also, 1445 cm$^{-1}$ is characteristic of the amide III C-N stretching vibrations. These peaks confirm to presence of urea in CNPsu. Also, Activation CNPsu in nitrogen furance at 900° C was caused to create N-H groups and C≡N in wave numbers of 3438 cm$^{-1}$ and 2046 cm$^{-1}$, respectively. Moreover, the intensity of peaks in wave number 2921 cm$^{-1}$ reduced to create -CH$_3$ groups due to decrease hydrogen groups with activation CNPs and CNPsu in high temperature. On the other hand, they have C≡C group in the abosprtion bond at 2046 cm$^{-1}$ to illustrate the removal hydroxyl groups in them that are shown in Fig. 1(b). Wu et al. reported similar results about the presence of N-H and C-N groups in CNPsu with a broad absorption around 3100 cm$^{-1}$ and 1405 cm$^{-1}$, respectively [4]. Also, Zhao et al. studied functional groups of N-CQDs with peaks at 3265 cm$^{-1}$, and 1429 cm$^{-1}$, which are assigned to stretching vibrations of N-H/O-H, and stretching vibration of C-N, respectively [10].

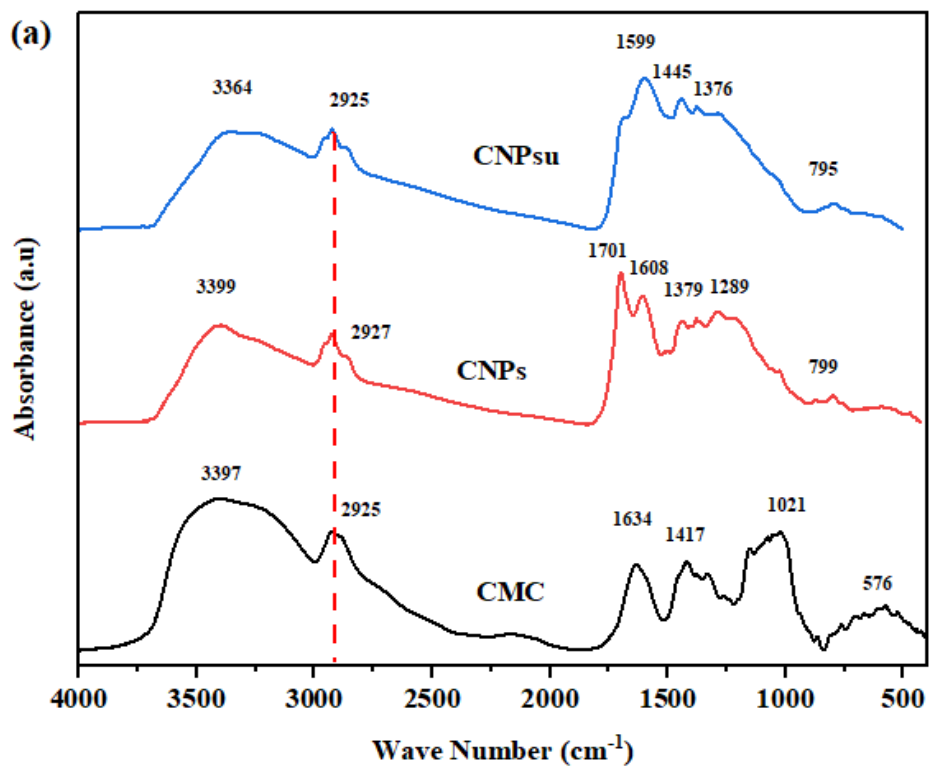

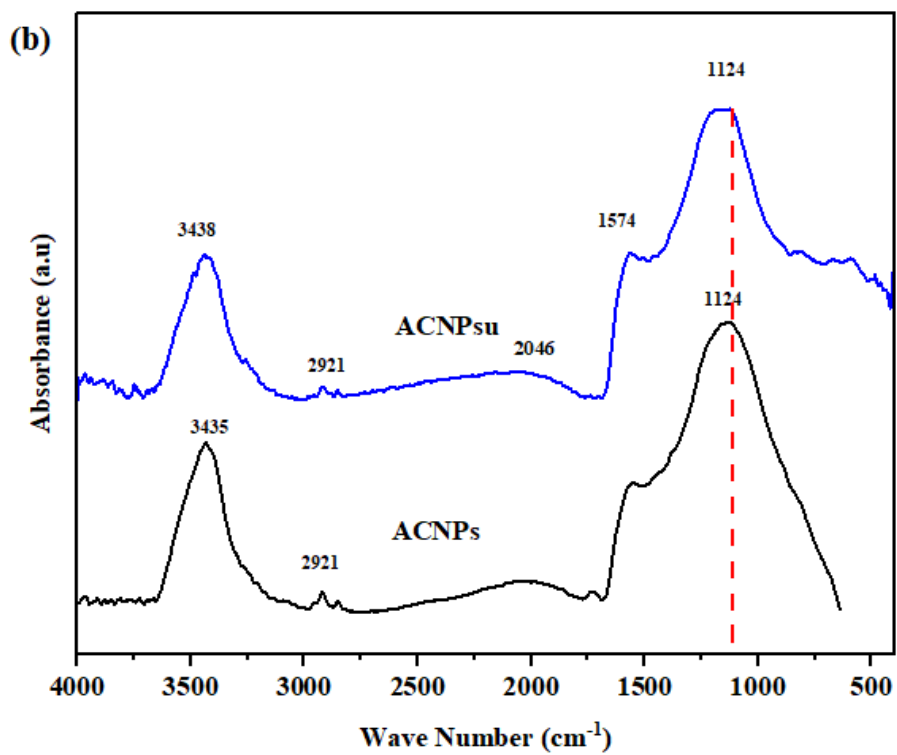

**Fig. 1.** FTIR images a) CMC, CNPs and CNPsu, b) ACNPs and ACNPsu.

### 3.2. H-NMR (Nuclear Magnetic Resonance)

Proton Nuclear magnetic resonance spectroscopy (HNMR) characterization of CNPs, CNPsu, ACNPs, and ACNPsu was performed in Acetone solvent that are shown in Fig. 2. Analysis of HNMR with MestReNova software for samples follows:

CNPs:

HNMR (400 MHz, Acetone-*d*) $\delta$ 2.98 (s, 2H), 2.13-2.04 (m, 18H), 1.30 (s, 1H).

CNPsu:

HNMR (400 MHz, Acetone-*d*) $\delta$ 2.98 (d, J= 10Hz, 1H), 2.92-2.85 (m, 4H), 2.15-1.94 (m, 2H), 1.30 (s, 0H).

ACNPs:

HNMR (400 MHz, Acetone-*d*) $\delta$ 2.99 (s, 3H), 2.18 (s, 6H), 2.15-2.07 (m, 4H), 2.02 (t, J=22 Hz, 0H), 1.30 (s, 0H).

ACNPsu:

HNMR (400 MHz, Acetone-*d*) $\delta$ 2.89 (s, 1H), 2.15-1.98 (m, 2H), 1.99 (s, 1H), 1.30 (s, 0H).

According to HNMR analysis the most significant signals in the range of 1-4 ppm for aliphatic protons, which belonged to C-H stretching and various hydrogen of aliphatic carbons that were attached with amine groups in their structure. Kumar et al. investigated HNMR carbon quantum dots (CQDs) with most significant signals in the range of 6-8 ppm that belonged to the hydrogen of aromatic rings and 1-4 ppm assigned to aliphatic protons [13]. Seo et al. evaluated chemical shifts of CQDs with peaks between 2.5 and 5 that were categorized as hydrophilic molecules, while, other peaks belonged to hydrophobic groups [14].

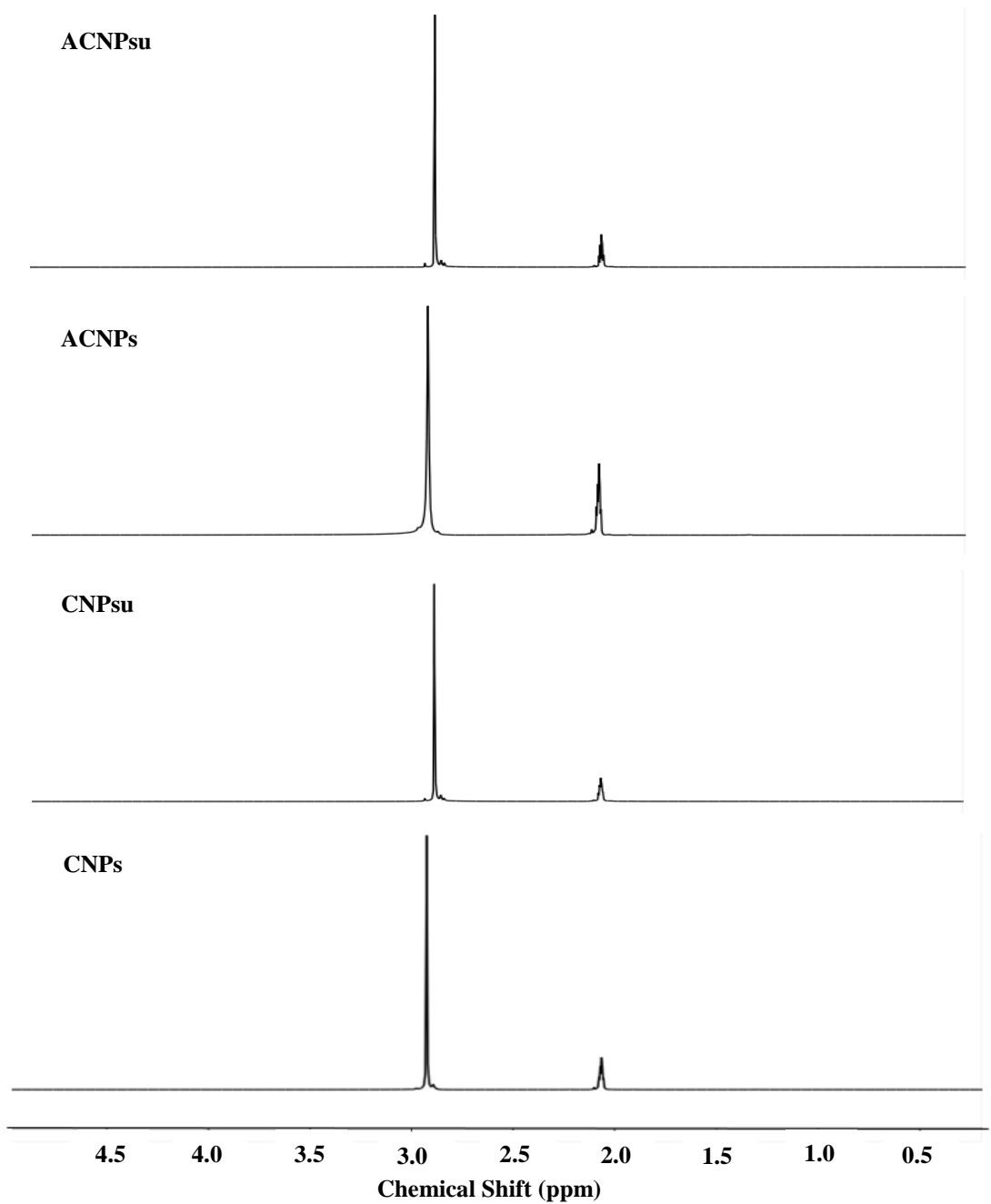

**Fig. 2.** H Nuclear magnetic resonance spectra of CNPs, CNPsu, ACNPs, ACNPsu.

3.3. Energy-dispersive X-ray spectroscopy (EDX)

Investigation of the purity and analytical spectrum of the percentage of elements in the CNPs, CNPsu, ACNPs, and ACNPsu was performed using EDX analysis. In the CNPs sample, two elements including carbon, and oxygen were observed with 72.80% and 25.84% weight, respectively. C and O elements in the CNPs are due to the sample structure during its synthesis by HTC method. The presence of nitrogen with 1.35% weight in CNPs is because of other compounds in the sample such as bacterial residues and other macromolecules in the bacterial extract during CNPs synthesis (Fig. 3(a)). The incorporation of urea in CMC solutions in the formation of CNPsu resulted to increase in efficacy of the synthesis process. Therefore, the percentage of carbon and nitrogen increased in CNPsu compare to CNPs (Fig. 3(b)). However, the weight percentage of nitrogen in CNPsu increased by about 4.16% compared to CNPs due to the presence of urea in its structure, the amount of oxygen decreased because of the bonding between nitrogen groups with carbon and hydrogen groups. On the other hand, ACNPs and ACNPsu had oxygen weight percentages of approximately 3.71%, and 7.13% lower than CNPs and CNPsu, respectively, which are shown in Fig. 3(c), and Fig. 3(d). Furthermore, the percentage of carbon in the state of activation NPs was higher than other samples, because of the increase in aromatic rings and removing hydroxyl groups in their structures. Lafta et al. obtained similar results of different elements in synthesis CNPs including carbon, oxygen, and nitrogen with 42.51%, 34.32%, and 23.17% weight, respectively [2]. Also, Viswanathan et al. evaluated elements of CNPs with weight percentages of 77.25%, 20.4% including carbon and oxygen, respectively. The presence of oxygen in CNPs might be due to the carbonization process [15].

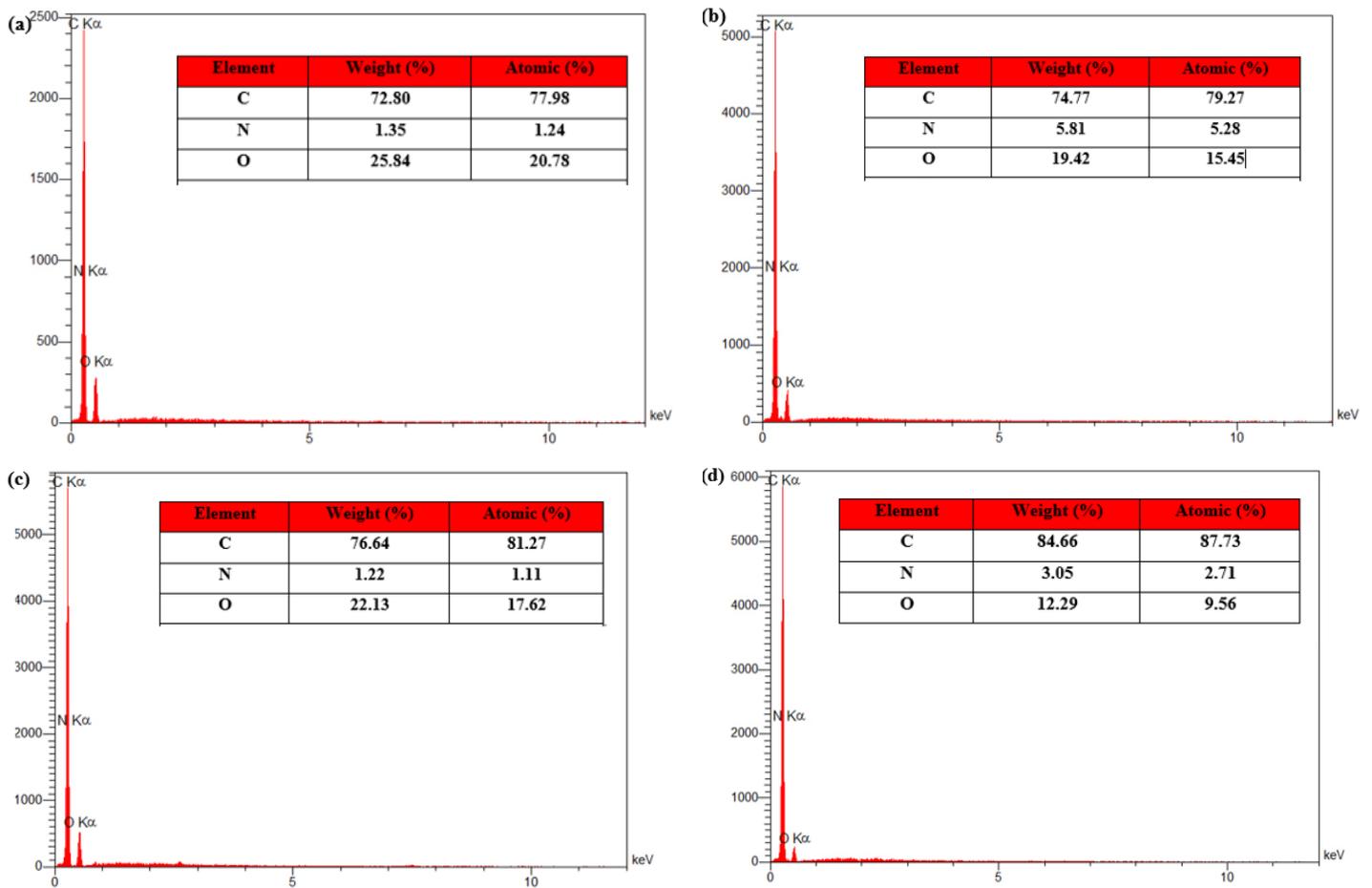

**Fig. 3.** EDX spectrum of a) CNPs, b) CNPsu, c) ACNPs, and d) ACNPsu.

### 3.4. Morphology of NPs

Morphology images of CNPs and ACNPs without urea and with incorporation of urea are shown in Fig. 4. Under all synthetic conditions, spheres were obtained. The morphology of CNPs without the addition of urea were spheres with a smooth surface and size distribution of particles approximately 2-4 µm that are demonstrated in Fig. 4(a). N-doping with the incorporation of urea in the CMC solution to synthesize CNPsu was triggered to significant adhesion of the neighboring spheres that can be observed with decreasing size distribution of particles about 1 µm (Fig. 4(b)). The main reason was a Maillard-type reaction, in which amines interact with the carbonyl group to reduce sugar molecules, resulting in the formation of a 1-deoxy-2-amino-1-ketone sugar. Depending on the system's pH, this reaction can follow various complex pathways, leading to the formation of nucleation seeds at different times and sizes. Incorporation of urea in 0.2 g with CMC solutions for the synthesis of CNPsu is similar to the results that another researhcer obtained with mixing urea and CMC solution to synthesize carbon dots. Also, the surface of the carbon spheres became rougher and bulker, indicating a surplus of urea [4]. Moreover, activation of CNPs and CNPsu in an nitrogen furnace at 900°C was caused to decrease the distribution size of particles to 1 µm and 400 - 600 nm, respectively (Fig. 4(c) and Fig. 4(d)). Because, activation of particles was triggered to aggregate and adhesion them due to van der Waals interaction between molecules in high temperatures and the erosion of carbon on the surface. On the other hand, TEM images show the morphology of ACNPsu that had the lowest size diameter distribution with spherical particles (Fig. 5). The similar results about morphology of CNPs by other researchers were obtained. Gairola et al. observed ACNPs have a smoother surface in compared to CNPs, with an average particle diameter of 74 nm [16]. Also, Wu et al. showed that activation of CNPsu was triggered to decrease diameter of carbon particles from 3-8 µm to 1-5 µm due to loss its impurities and the erosion of carbon on the surface [4].

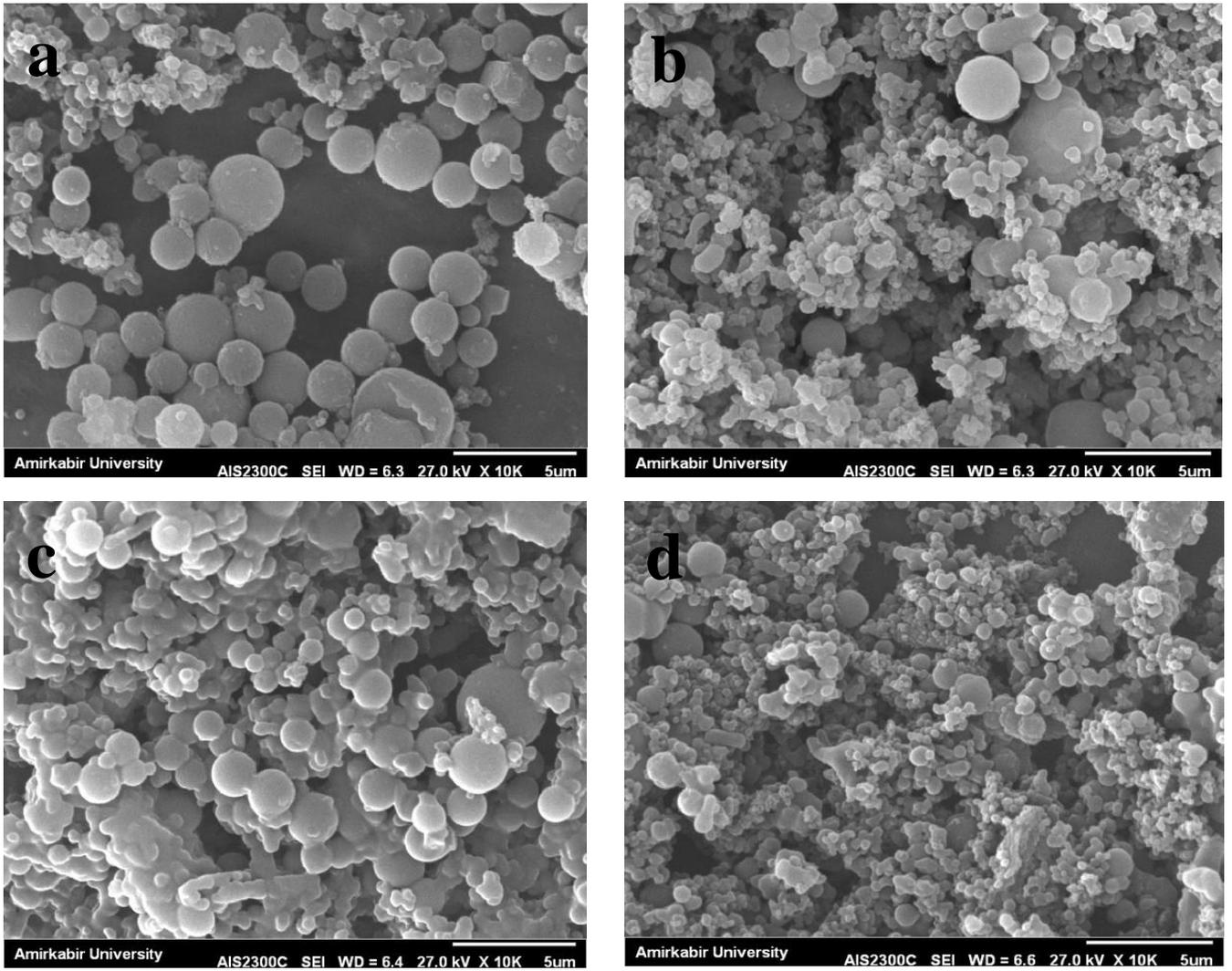

**Fig. 4.** SEM of nanoparticles a) CNPs, b) CNPsu, c) ACNPs, and d) ACNPsu.

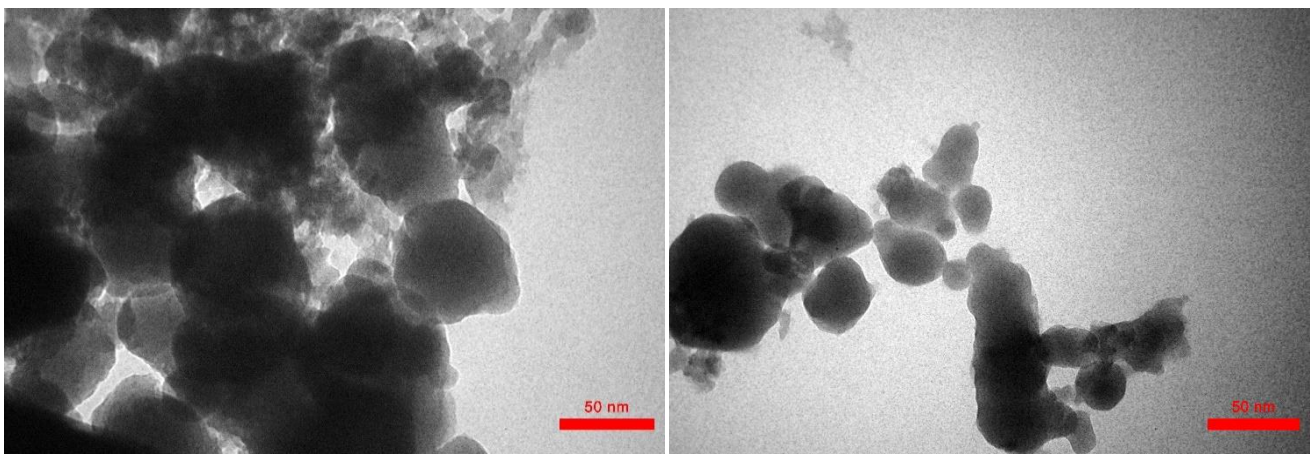

**Fig. 5.** TEM images of ACNPsu in 50 nm magnification.

### 3.5. Dynamic Light Scattering (DLS)

The size of particles, z-average and proton disulfide isomerase (PDI) have been measured using the DLS technique. If the particles have PDI < 0.1, the particle size distribution is so narrow. For the particles that have a uniformly distributed size, their PDI is between 0.1 and 0.7. While particles with PDI > 0.7 have a broad size distribution. Figure 6 illustrates distribution statistics for different types of NPs using the SBL algorithm. The results show CNPs with PDI of about 0.00546 with z-average 293 nm had a narrow size distribution. Thus, 82% of particles were in size of 187 nm and 17% of them had a size of about 472 nm that are shown in Fig. 6(a). The incorporation of urea into CMC solution to synthesize CNPsu resulted in a broad size distribution with a PDI of about 0.60533 which is so close to PDI~0.7 that are illustrated in Table 1. On the other hand, adding urea to produce CNPsu was triggered to decrease particle size by about 40 nm with a z-average of about 606 nm (Fig. 6(b)). Activation of CNPs and CNPsu in high temperatures at 900° C helped to reduce particle size with uniformly distributed size with PDI < 0.7. Thus, the particle size of ACNPs and ACNPsu was 28 nm and 51 nm, respectively. In this state, activation of NPs was caused by a considerable reduction in particle size which are indicated in Fig. 6(c) and Fig. 6(d). Moreover, adding urea to CMC solution for producing ACNPsu increased particle size in compared to ACNPs without urea. The main reason was related to the aggregate of particles by adding urea to the CMC solution and activation of them by producing triple bonds in their structure to interact with urea. Ray et al. with the investigation of the size distribution of carbon particles and carbon nanoparticles observed that CPs had a broad size distribution and larger sizes of about 20- 350 nm compared to CNPs that had a narrow size distribution with an average diameter of 12.5 nm [17]. Also, Chen et al. observed that the size distribution of CNPs was in the range of 18 to 70 nm [18].

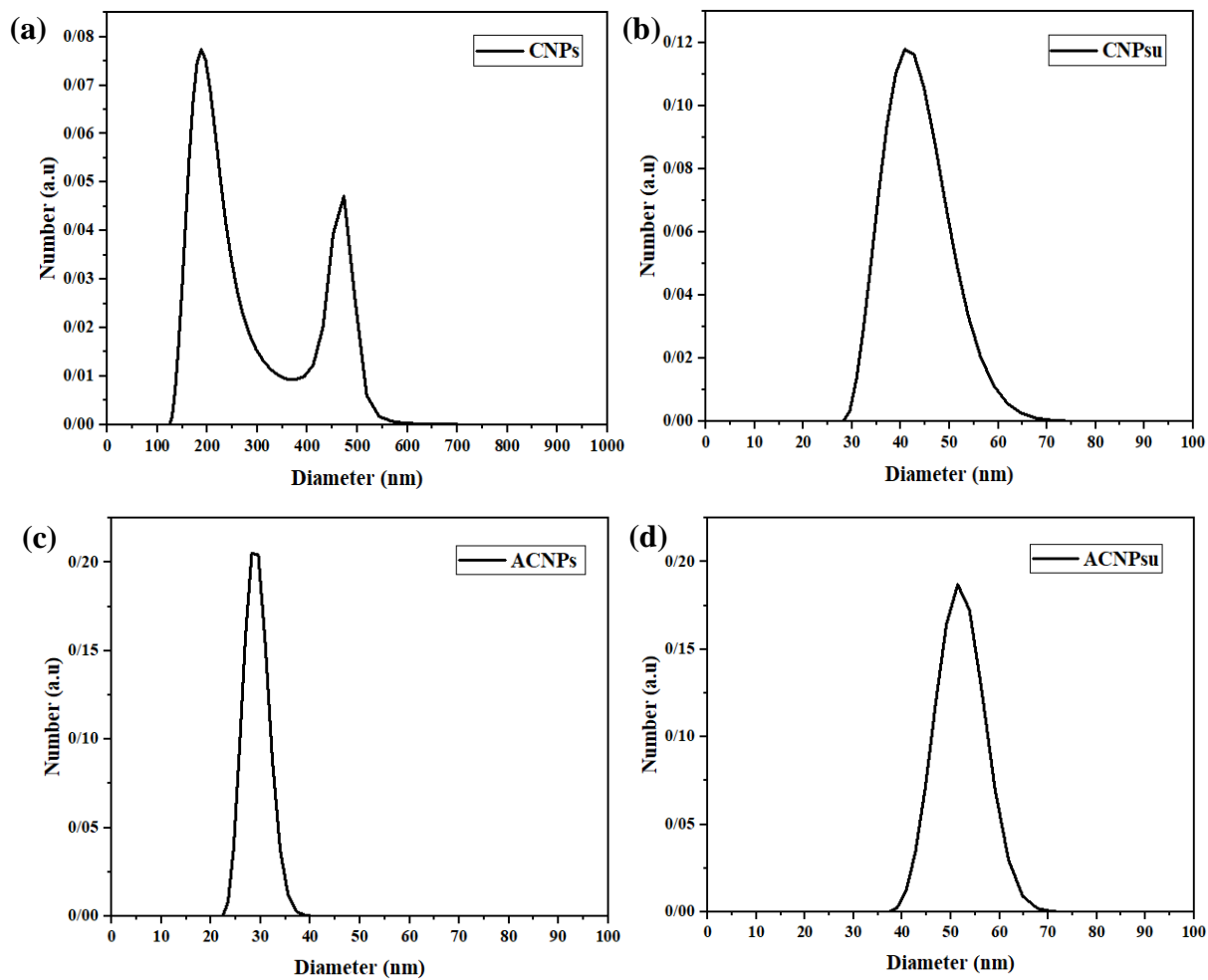

**Fig. 6.** Distribution statistics of Particle size according to SBL algorithm a) CNPs, b) CNPsu, c) ACNPs, d) ACNPsu.

Table 1. Characterizations of different types of NPs.

| Samples | PDI | Z-average (nm) | Particle size (nm) | Surface area (m$^2$/g) | Pore size (nm) | Pore volume (cm$^3$/g) |
|---|---|---|---|---|---|---|
| CNPs | 0.00546 | 293 | 187 | 0 | μ | - |
| CNPsu | 0.60533 | 606 | 40 | 41.7 | 84.69 | 0.088 |
| ACNPs | 0.22447 | 76 | 28 | 552.9 | 33.6 | 0.46 |
| ACNPsu | 0.32073 | 154 | 51 | 351.01 | 31.87 | 0.28 |

### 3.6. Brunauer-Emmett-Teller (BET)

The surface areas and textural properties for different types of NPs were determined using the BET method by the physical adsorption of $N_2$ gas molecules at 77 K. Before measurement, samples were degassed under vacuum in $N_2$ for 2 h at 150°C. Absorption isotherm of $N_2$ gas for three points of samples are shown in Fig. 7. The characterizations of samples in surface area, total pore size distribution, and total pore volume are shown in Table 1. Also, the nitrogen adsorption-desorption isotherms have been shown in Fig. 8. According to the figures and table, CNPs have micrometers of pore size without nitrogen adsorption-desorption isotherm. The calculated surface areas for CNPsu, ACNPs, and ACNPsu are 41.77 m$^2$/g, 552.9 m$^2$/g, and 351.01 m$^2$/g, respectively (Table 1). However, the Addition of urea to CNPs was caused to increase absorption isotherm with a reducing average pore size of about 84.69 nm, and activation CNPsu was triggered to reduce surface area to absorb $N_2$ gas rather than ACNPs due to the aggregate of particles. While, ACNPsu have average pore size, and total pore volume about 31.87 nm and 0.28 cm$^3$/g lower than ACNPs (Table 1). Thus, total pore volume and total pore size with loading urea and activation CNPs reduced due to a decrease in their diameter and increased surface area. Similar results were obtained in DLS and SEM. The isotherm of the CNPsu (Fig. 8(a)) exhibited a combination of Type I and Type IV isotherms, with a broad distribution of pore sizes, and significant peaks in both the micropore and mesopore ranges while maintaining a high surface area. The slight reduction in surface area of CNPsu compared to the activated carbon. While, ACNPs, ACNPsu exhibited a type I isotherm (Fig. 8(b), and Fig. 8(c)), indicative microporous materials with a high surface area and significant adsorption at low relative pressures. The high surface area was characteristic of activated carbons with their high adsorption capacity. While, the Barrett-Joyner-Halenda (BJH) pore size distribution analysis confirmed the presence of nanoporous with a dominant peak at 90° A, 62° A, and 47° A in CNPs, ACNPs, and ACNPsu, respectively that are shown in Fig. 9. Alsheheri et al. analyzed nitrogen adsorption-desorption behavior, identifying type I isotherms in microporous activated carbon particles, characterized by high surface area and strong adsorption at low relative pressures. BJH micropore distribution, with a dominant peak at 17.4 Å and a surface area of 1102 m²/g. Incorporation of alumina into activated carbon, which resulted in a Type I & Type IV isotherm combination, due to integration of alumina particles, yielding a surface area of 778 m²/g and a pore diameter of 23.7 Å [19]. Moreover, similar results showed that surface area and the pore volume for ACNPs were lower than CNPs because of microwave-induced rapid heating, which could trigger smaller nanoparticles to coalesce into larger structures [16].

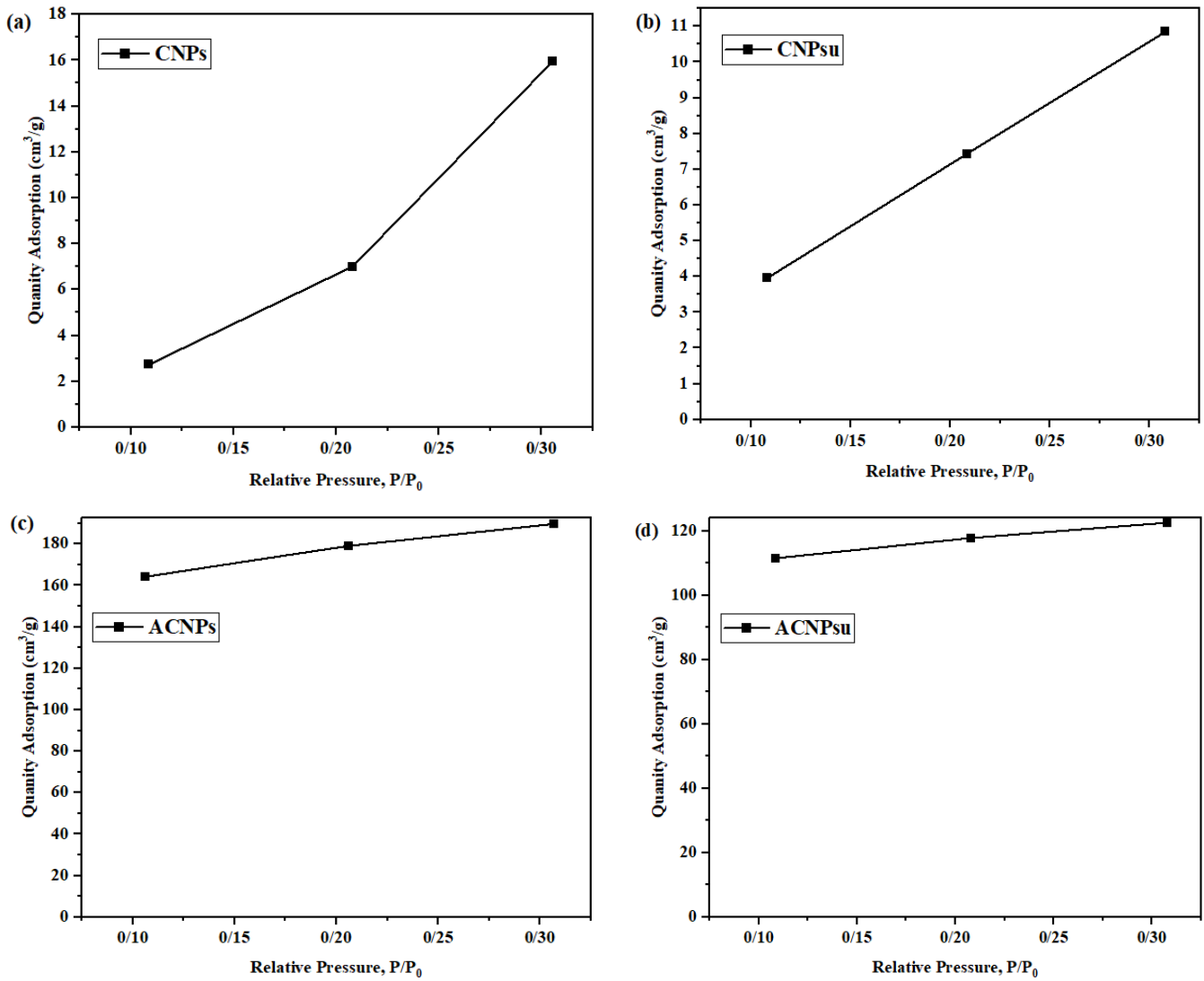

**Fig. 7.** Quanity absorption of samples in relative pressure for three points, a) CNPs, b) CNPsu, c) ACNPs, and d) ACNPsu.

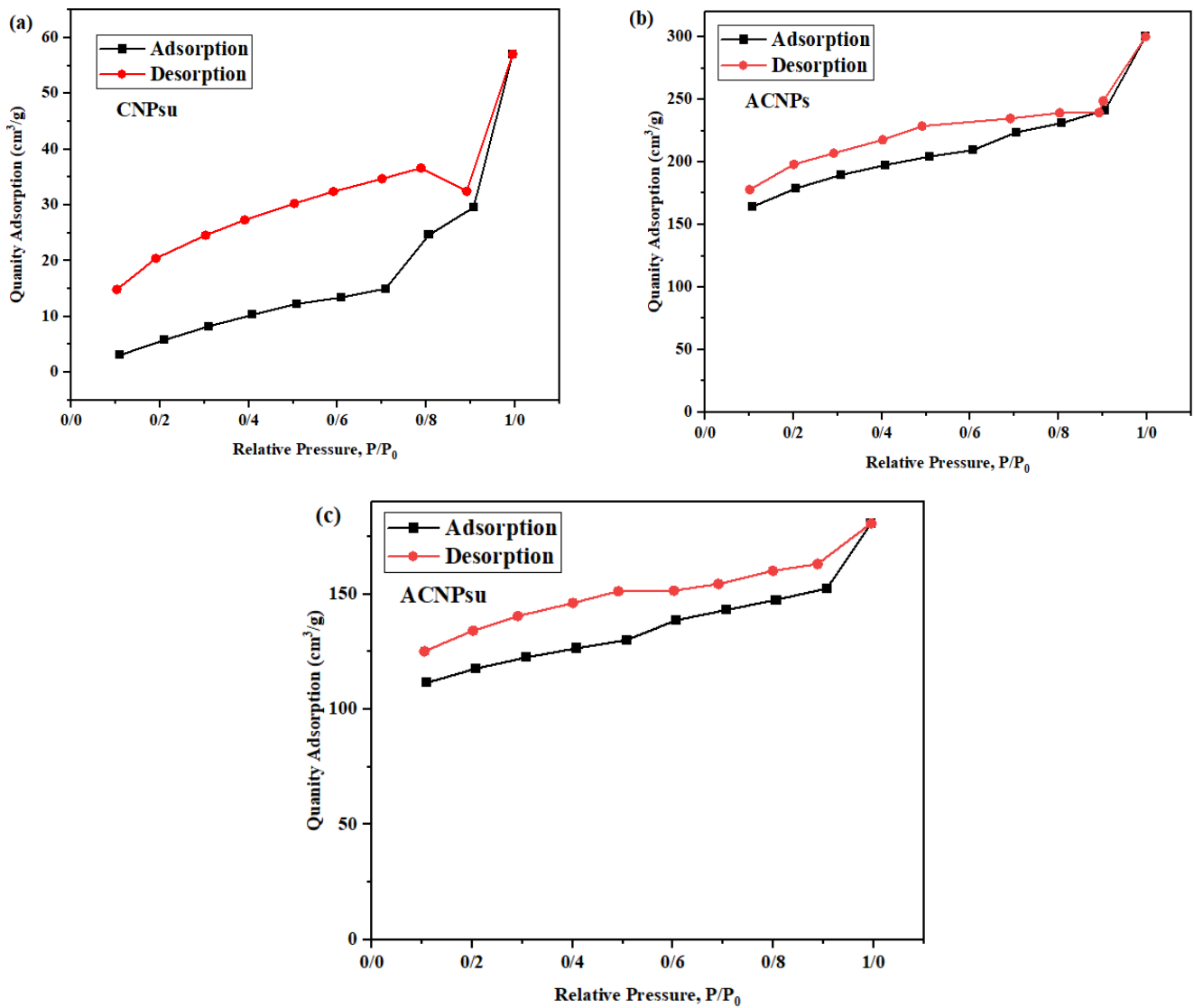

**Fig. 8.** Nitrogen adsorption/desorption isotherm (BET) of a) CNPsu, b) ACNPs, and c) ACNPsu.

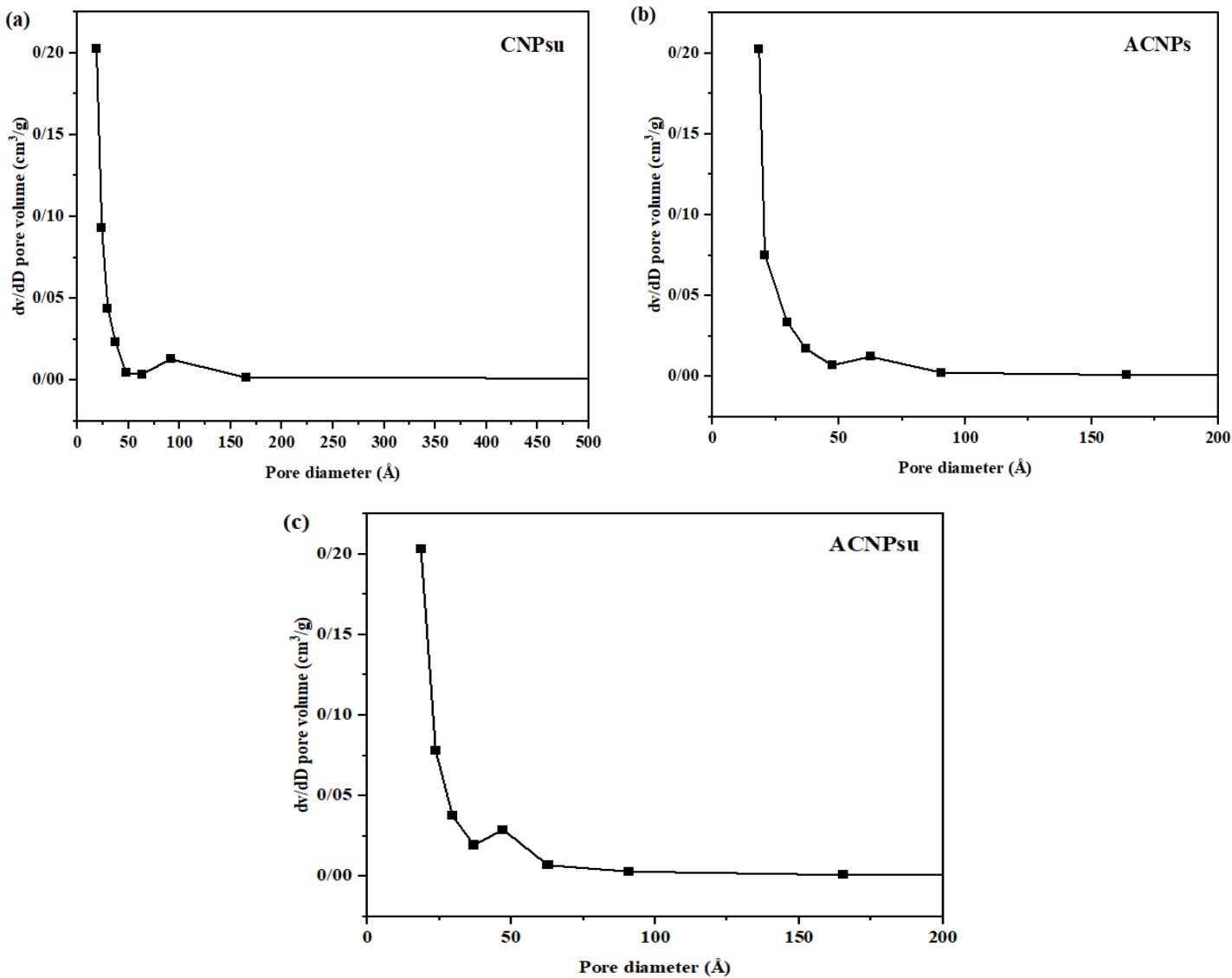

**Fig. 9.** BJH adsorption pore size distribution a) CNPsu, b) ACNPs, and c) ACNPsu.

3.7. Thermo gravimetric analysis (TGA)

TGA curves for CNPs, CNPsu, ACNPs, and ACNPsu that were synthesized from CMC/urea solutions are shown in Fig. 10. This diagram shows weight reduction in the samples with increasing temperature. The temperature raised from room temperature to 700° C with a heating rate of 10°C per minute in nitrogen atmosphere. TGA profile for samples showed that moisture was removed from samples up to 100°C. Significant weigh loss did not occurred up to 250°C for CNPs and CNPsu and up to 450°C for activated samples. Also, CNPs and CNPsu experienced sharp weight loss, indicative of structural degradation, which occurred around from 280°C to 700°C and from 350°C to 480°C, respectively. The presence of urea in CNPsu increased carbon particle efficacy, delaying weight loss until higher temperatures. On the other hand, activating NPs resulted in thermal stability in high temperatures up to 450°C (ACNPS and ACNPsu samples). This enhanced stability is attributed to increase aromatic rings formation during activation. Also, ACNPsu exhibited weight loss with slow slope rather than ACNPs between 450°C and 500°C, due to the incorporation of urea in their structure that was caused to augment the amount of carbon with increasing efficiency in their synthesis. Thus, degradation occured in the structure of samples with increasing temperature was due to the decomposition of hemicellulose, cellulose, and lignin in their structure. Moreover, the percentage of weight loss for samples CNPs, CNPsu, ACNPs, and ACNPsu at 700°C was approximately 68%, 55%, 39%, and 32%, respectively. Therefore, activation of NPs and incorporation of urea increase structure stability at high temperature compared to other samples. Poongavanam et al. obtained similar results that belonged to activated carbon from Kigelia Africana leaveas with major weight loss about 40% from 50° C to 910° C in two stages at 67.3° C-500° C, and 500° C-910°C, due to degradate of material strucutres [20].

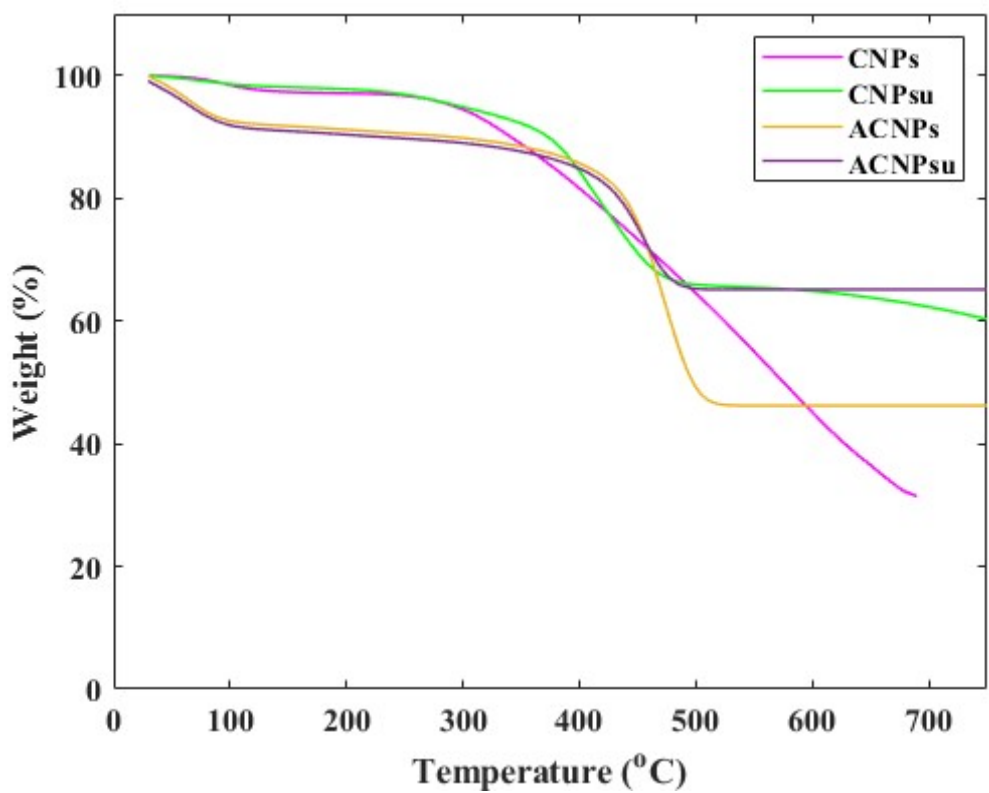

**Fig. 10.** TGA curve of CNPs, CNPsu, ACNPs, and ACNPsu.

3.8. Zeta Potential

The electrical properties and particle charge on their surface in solution were determined by zeta potential. The zeta potential and electrophoretic mobility mean were evaluated at approximately -32.1 mV and – 0.000166 cm$^2$/Vs, respectively (Fig. 11). The negative charge in carbon nanoparticles is due to carboxyl groups in the CMC solution to synthesize carbon black. Similar results observed with evaluation zeta potential of carbon particles approximately -20.4 mV and -28.2 mV. This amount indicates that the surface of the particles is naturally acidic. Also, the range of pH and surfactant are important factors in studying the particles' charge [18, 21].

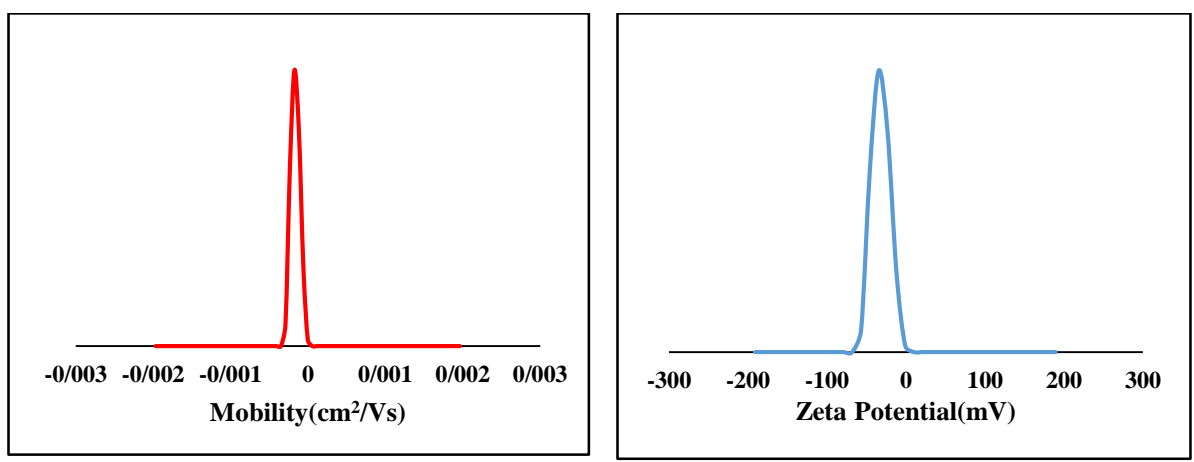

**Fig. 11.** Zeta potential and mobility of ACNPsu.

3.9. Encapsulation efficiency of drug

The encapsulation efficiency of CD was studied in four types of carbon nanoparticles for 24 h. Different parameters such as nanoparticle size, drug concentration, activated carbon nanoparticles, and incorporation of urea to synthesize NPs have been affected on CD encapsulation into them. The incorporation of urea with CMC solution to synthesize CNPsu in comparison to CNPs without urea was caused to decrease in pore size and an increase surface area of NPs significantly. Thus, a drastic decrease in particle size with increasing surface area and augment in drug concentration were caused to increase in CD entrapment into NPs. The results illustrated that increasing CD concentration from 0.0005 g/mL to 0.002 g/mL for loading into NPs with a concentration of 0.001 g/mL increased the percentage of drug entrapment from 39.27% to 82.2% that are shown in Fig. 12. According to the results of DLS, BET, and SEM, activation NPs were triggered to increase considerable surface area and decrease pore size. Thus, ACNPs had drug entrapment percentages more than CNPs in different concentrations of CD. So, loading 0.001 g/mL CD into CNPsu, and ACNPs was caused to increase in drug entrapment from 51.68% to 84.69%, respectively. On the other hand, entrapment of 0.002 g/mL CD into CNPsu was higher than ACNPs with 79.17% and 54.6%, respectively. Because activation of NPs was caused to aggregate of them, and with increasing concentration of drug, a further part of CD was placed on the surface of particles. Hence, more amount of the drug could be dissolved in the PBS. Furthermore, ACNPsu in high concentration of drug had high drug entrapment rather than to ACNPs due to the incorporation of urea in the structure of ACNPsu to increase surface area. Mandegari et al. investigated the amount of CD in PVA nanoparticles at different times. The results illustrated that increasing the content of the drug from 5% to 10% w/w was caused to adecrease drug entrapment efficiency from 87.5% to 74.6% due to more exposure of the drug to the water [22].

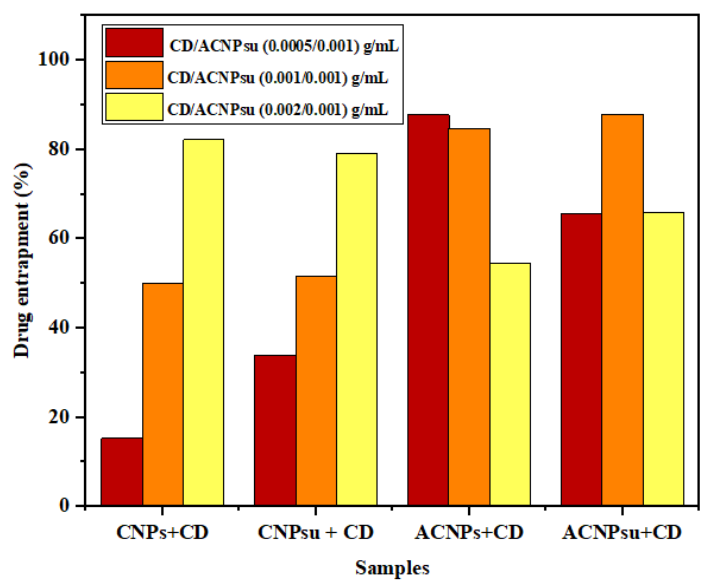

**Fig. 12.** Effect of drug concentration on the drug entrapment into ACNPsu during 24 h.

## 4. Conclusion

In this work, different types of NPs including CNPs, CNPsu, ACNPs, and ACNPsu were synthesized using one-pot HTC method from carboxymethyl cellulose (CMC) and urea. The results showed that N-doping in the structure of CNPs and activation of them in high temperatures were triggered by significant adhesion of the neighboring sphere due to van der Waals's interaction between molecules in high temperatures and the erosion of carbon on the surface. Also, ACNPsu had the lowest size distribution with spherical particles and a particle size of approximately 51 nm. However, activation of CNPsu was caused to increase particle size in comparison to ACNPs due to the aggregate of particles with N-doping and producing triple bonds in their structure to interact with urea. Moreover, activation of NPs with urea in high temperatures was caused to generate different functional groups including N-H, C-N, C≡N. On the other hand, the chemical structure of NPs showed that activation of N-doping

NPs was triggered to increase in aromatic rings due to augment in the percentage of carbon and removal hydroxyl groups in their structure. Also, increasing in aromatic rings in the structure of NPs with activation of them was caused to high temperature stability in compared to other samples. Another characterization of the activation of NPs was high surface area with their high adsorption capacity for drug delivery application. Therefore, ACNPsu with a negative charge of about -32.1 mV due to the incorporation of urea in their structure to increase surface area had a high encapsulation efficiency of approximately 84.69% in 0.001 g/mL CD with a positive charged drug.